# Absorption control in pseudo disordered photonic-crystal thin films

Romain Peretti, Guillaume Gomard[†], Loïc Lalouat, Christian Seassal, and Emmanuel Drouard*

*Institut des Nanotechnologies de Lyon (INL), Université de Lyon, UMR 5270, CNRS-INSA-ECL-UCBL*
*Ecole Centrale de Lyon, 36 Avenue Guy de Collongue, 69134 Ecully Cedex, France*

*† Present address: Light Technology Institute (LTI), Karlsruhe Institute of Technology (KIT), 76131 Karlsruhe, Germany*

**Corresponding author: emmanuel.drouard@ec-lyon.fr*

**Abstract**: The positive effects of various perturbations introduced in a bidimensional photonic crystal patterned membrane on its integrated absorption are investigated numerically and theoretically. Two phenomena responsible for the enhanced absorption observed are identified: an increase of the spectral density of modes, obtained thanks to folding mechanisms in the reciprocal lattice, and a better coupling of the modes with the incident light. By introducing a proper pseudo-disordered pattern, we show that those two effects can be exploited so as to overcome the integrated absorption obtained for an optimized and single pattern unit cell Photonic Crystal.

## I. INTRODUCTION

Light trapping is a topic of prime importance for many fields of engineering and science [1]. More precisely, controlling the absorption of the light is essential for applications such as solar [2] and indoor photovoltaics [3] or sensors [4]. The general trend towards cost reduction leads the photovoltaics community to use (ultra)thin layers of active materials, with thicknesses ranging from a few hundreds of nanometers to a few micrometers, depending on the extinction coefficient of the material. Unfortunately, this thickness decrease comes along with an incomplete absorption of the light, especially close to the energy bandgap of the active media, leading to a low cell efficiency.

That is why Photonic Crystals (PhC) were introduced in such ultrathin film devices, in order to add resonant modes to enhance the absorption. Theoretically, such a patterning enables reaching an absorption at a given wavelength as high as 100% [5, 6]. To go further on photovoltaic-applied photonic concepts, two main points have to be satisfied: to get as many modes as possible per wavelength unit [7, 8], and to make those modes absorb as much as possible [9].

If the second point can be reached using PhCs, the first one is really difficult to obtain using a PhC wherein the unit cell is made of a single pattern. Such a PhC membrane does not exhibit enough modes to be the most effective structure on a wide spectral range as needed for solar cells. It has thus been proposed to use more complex patterns, such as multiperiodic [8], pseudo-disordered [10, 11, 12, 13] (i.e., a disordered cell periodically repeated), or disordered [14, 15] structures. The resulting enhanced optical absorption was attributed either to the broadening of the absorption peaks [16], or to the modification of the spatial Fourier spectrum [13] thanks to the more complex pattern leading to additional modes which have the potential to couple with the incident light. However, those studies did not propose a thorough description of the origin and of the properties of those new modes responsible for the absorption increase. Moreover, there is no comparison between the integrated absorptions of these designs and that of an optimized single pattern PhC.

In this paper, we aim at explaining how super cells, composed of different perturbed patterns, enable increasing the absorption of single period photonic structures. Indeed, such super cells are expected to allow a more efficient coupling into Bloch modes, as well as an increase of the density of modes, even in comparison to an optimized single pattern unit cell PhC.

To achieve this, we use a perturbative method, by first introducing multiple pattern super cells in PhC, with a few modified patterns compared to the single pattern unit cell PhC, and controlled symmetry properties. Then, we implement pseudo-disordered PhC, i.e. super cells where most of the properties of the pattern are determined using a statistical distribution.

All these structures are depicted in section 2. Both numerical and phenomenological approaches, described in section 3, are used to describe the absorption properties of the structures. The results are reported in section 4 for simple cases. In section 5, an in-depth analysis of the results is provided thanks to the analysis of the single and multiple pattern structures in the reciprocal lattice space. The integrated absorption enhancement of a multiple pattern structure compared to an optimized single pattern structure is finally computed in section 6.

# II. DESCRIPTION OF THE PATTERNED MEMBRANES

## A. Absorbing membrane structured as a single pattern unit cell PhC

The present study relies on a simple, easily feasible structure made of a 195 nm thin hydrogenated amorphous silicon (a-Si:H) membrane lying on a silica substrate. A 2D square lattice of nano-holes is fully etched in the a-Si:H layer, as shown in Fig.1.

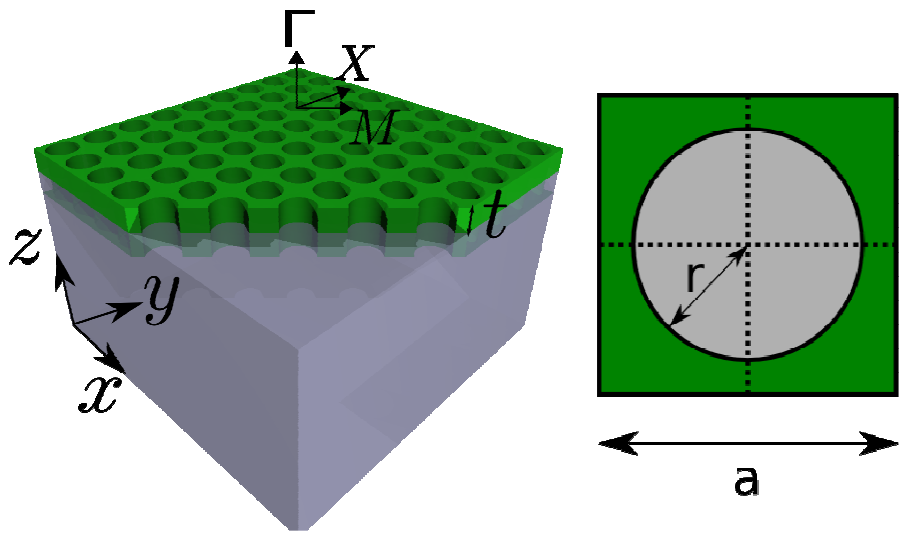

Fig.1 : (Left) Patterned a-Si:H layer ($C_{4v}$ symmetry) on its silica substrate, with in-plane (X, M) and vertical (Γ) crystallographic directions of the structure. (Right) Top view of a single pattern unit cell, with a period *a* and a radius *r*

The resulting single pattern 2D PhC membrane is shown in Fig.1. It is defined by its period *a*, and its surfacic air filling fraction, $ff = \pi r^2/a^2$, *r* being the radius of the hole, which is set to 50% in this work. Indeed, this value can be easily obtained with large-area fabrication techniques and is known to lead to a high absorption [17].

## B. Multiple pattern super cell PhC

To increase the integrated absorption, new addressable modes have to be introduced, possibly thanks to a periodic perturbation in the previous single pattern unit cell lattice. For the sake of simplicity, we chose to simulate and analyze such perturbations by introducing a super cell consisting of 3 *x* 3 patterns having a whole size set as 3*a x* 3*a*. In this framework, a perturbation can be introduced in the super cell by moving the holes (scaling the holes or changing their shape will not be considered in the scope of the study), leading to a constant value of *ff*. Depending on the perturbation, the 3 *x* 3 super cell with a multiple pattern can either keep the original $C_{4v}$ point group symmetry operations [18] or loose some or all of the symmetry operations of the point group.

To maintain all the original $C_{4v}$ symmetry operations, only two kinds of perturbations are then allowed. One can: **1** -move the centres of the 4 holes located in the corners of the super cell along its diagonals, or **2**- move the centres of the 4 holes at the middle of the sides towards the centre of the super cell. These perturbations will be referred to as symmetric perturbations.

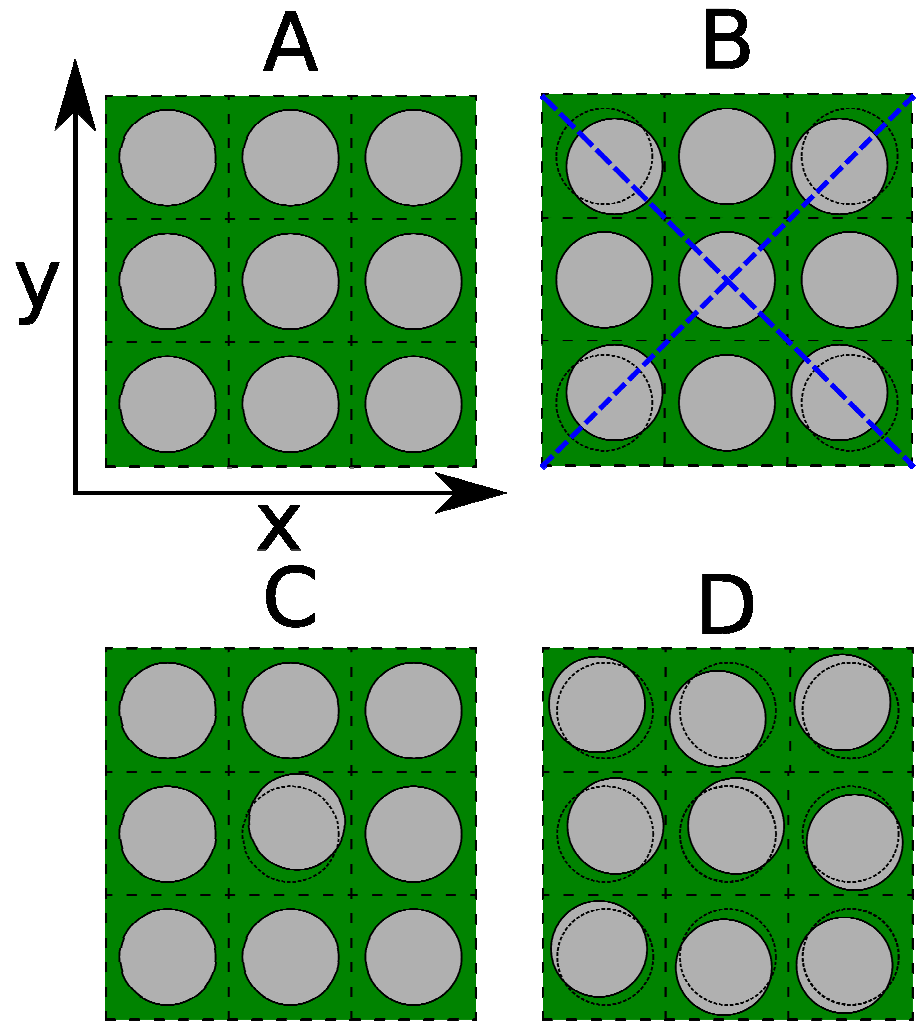

Fig.2 : 3*a x* 3*a* 2D PhC. A: Single pattern unit cell PhC used as a reference , B to D: Multiple pattern super cells with symmetric perturbation (diagonals directions represented by the blue dashed lines) (B), with non-symmetric perturbation (shift along 60° angle) (C),and with pseudo-disordered perturbation (D). The contours of the unperturbed holes are depicted with the black dashed lines.

All the investigated perturbations are shown in Fig.2.

The selected perturbations for this study consist in:

- Moving the four holes in the corner of the super cell towards its centre along the diagonals (symmetric case, i.e. satisfying all the symmetry operations of the $C_{4v}$ point group, Fig.2B)
- Changing the position of the central hole along a direction making a 60° angle with respect to the *x* axis (non-symmetric case, i.e. satisfying none of the symmetry operations of the $C_{4v}$ point group except identity, Fig.2C)
- Introducing disorder in the position of each hole which are moved within their own sub cell according to a normal distribution (pseudo-disordered case, satisfying none of the symmetry operations of the $C_{4v}$ point group except identity, Fig.2D)

Whatever the case, it is important to set the magnitude of this perturbation of the index profile. To draw comparisons between the different kinds of perturbation, a quantitative measurement of the magnitude of the opto-geometric perturbation, denoted *M*, can be defined as:

$$M(\lambda)=\frac{1}{a^2}\iint_{period}\left|n_{mp}(\lambda)-n_{sp}(\lambda)\right|dS \quad (1)$$

where nsp is the refractive index of the structure at the position S without perturbation (single pattern unit cell), nmp is the refractive index of the structure at the position S with perturbation (multiple pattern super cell), dS is the infinitesimal displacement area around the position S, and the surface is integrated along the x and y directions defined on Fig.1.

Considering our structure, one can define:

$$\Delta n(\lambda)=\left|n_{a-Si:H}(\lambda)-1\right| \quad (2)$$

equation (1) can then be written as:

$$M(\lambda)=\Delta n(\lambda)\left|\left(\frac{1}{a^2}\iint_{perturbation}dS\right)\right|$$
$$M(\lambda)=\Delta n(\lambda)\frac{\iint_{perturbation}\left|dS\right|}{a^2}=\Delta n(\lambda)\Delta S \quad (3)$$

where the first term consists in the index contrast and the second term is the surface where the index is changed because of the perturbation. Indeed, in our case, because the index change is always identical, the magnitude of the surface perturbation in square period units (ΔS) will be the most straightforward way to express the magnitude of the opto-geometric perturbation. Moreover, the surface perturbation does not depend on the wavelength.

In addition, we also analyse larger super cells in the sole case of the symmetric perturbation (as it is the less demanding in terms of computational resources) for 6 *x* 6 and 9 *x* 9 super cells. In these cases, instead of moving one hole at each corner, we move respectively 2 by 2 holes and 3 by 3 holes using the same displacement vector. This leads to the same magnitude of perturbation than for the 3 *x* 3 super cell.

# III. SIMULATIONS METHODOLOGY

The absorption values have been calculated by Finite-difference time-domain (FDTD) simulations [19] taking into account the chromatic dispersion of the optical indices of both a-Si:H and $SiO_2$ [20]. The agreement between ellipsometric measurements for a-Si:H and the fit used in the FDTD simulations (see Fig.14 in Appendix), enables trustworthy numerical simulations. Details about FDTD simulations can be found in Appendix.

The refractive index $n_{a\text{-}Si:H}(\lambda)$ and extinction coefficient $k_{a\text{-}Si:H}(\lambda)$ also enable to estimate the critical quality factor $Q_c$ of the mode, which maximizes its peak absorption [21]:

$$Q_c(\lambda)=\frac{n(\lambda)}{2k(\lambda)}\approx\frac{(1-ff)n_{a-Si:H}(\lambda)+ff}{2(1-ff)k_{a-Si:H}(\lambda)} \quad (4)$$

where n(λ) and k(λ) are the effective refractive index and extinction coefficient of a given mode in the structure. The second term in equation (4) is an approximation, since the rigorous formulation would require the preliminary computation of the complete electric field distribution.

Given an arbitrary mode, at the critical coupling, the external losses equal the absorption losses, enabling an optimal absorption at the resonant wavelength. It is also an upper limit to maximize the broadband integrated absorption [9]. An accurate numerical estimate of the Q factor of the modes involved in the absorption mechanism is achieved using a harmonic inversion algorithm (Harminv [22]). Comparing these Q factors to Qc enables a phenomenological analysis of the FDTD absorption results.

To highlight more clearly the observed effects, the integrated absorption under AM1.5G illumination [23] and the resulting short circuit current density Jsc (using a theoretical internal quantum efficiency of 1) are computed. It can be noticed that if the light is fully absorbed in the range of 400-750 nm, the highest achievable Jsc is 19.9 $mA/cm^2$.

# IV. ABSORPTION EFFICIENCY IN PERTURBED PHOTONIC CRYSTAL MEMBRANES

## A. Single mode, single pattern unit cell PhC membrane

The period *a* of the single pattern unit cell PhC membrane as shown in Fig.1 is set to 220 nm. With the previously set *ff*, this PhC membrane exhibits a single Slow Bloch mode (SBM) that can be excited by a plane wave under normal incidence (symmetric with respect to the in-plane crystallographic directions). The absorption of this PhC membrane is compared to the one of an equivalent but unpatterned stack in Fig.3 between 400 and 750 nm.

Equation (4) is also plotted in Fig.3 for *ff* = 50%. This enables to determine the spectral range where resonant modes (with Q>10) are needed to enhance the absorption, that is from 600 nm to 750 nm. We will therefore restrict ourselves to this wavelength range in the following. We then observe that the SBM of interest (at around 666 nm) contributes to the enhancement of the absorption [24], whereas the two other anti-symmetric SBM exist but cannot be excited by the vertically incident plane wave.

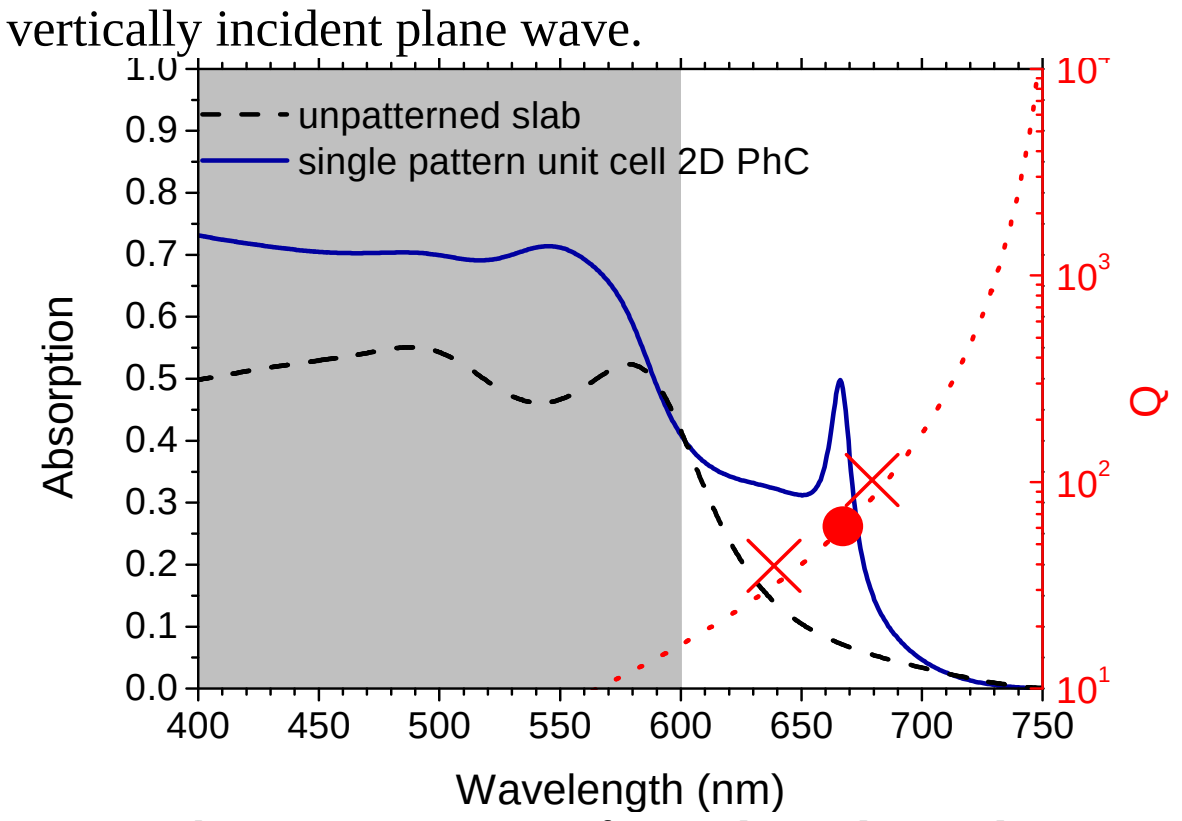

Fig.3 : Absorption spectrum of a single mode, single pattern unit cell 2D PhC membrane (*a*=220 nm, *ff*=50%) in blue, compared to the one of the non structured layer in dashed black. In red dotted line, with scale on the right, $Q_c$ calculated from equation (4), in red dot the *Q* of the mode at around 666 nm, in red crosses the *Q* of the non coupling modes found by dipole excitation simulations. The spectral range of interest lies above 600 nm.

Even if not optimized for maximizing the integrated absorption, this single mode PhC membrane can be used in the following to understand the effect of the introduced perturbations in the pattern.

## B. Multiple pattern super cell PhC membrane

In this subsection, we present the simulation results obtained after introducing the perturbations depicted in Fig.2.

Fig.4 and Fig.5 gather the absorption spectra at normal incidence for increasing magnitudes of surface perturbation ΔS (from top to bottom) for the three previously perturbations considered. For each case, the upper ΔS limit corresponds to a physical limit, where two holes are in contact. On the same figure, Q factors of the modes are reported and have been calculated thanks to their full width at half maximum around their peak wavelength (horizontal bar). The red dashed line represents the critical coupling value plotted in the wavelength range of interest (from equation (4)). Finally, because of the similarity between x and y polarizations, only the x polarization is presented for the non-symmetric and pseudo-disordered cases.

One can observe several modifications related to the introduction of the symmetric perturbation. First, when a small perturbation is introduced, new modes appear, resulting in additional peaks in the absorption spectrum. This occurs especially at large wavelengths, i.e. close to the bandgap of the material, where the absorption tends to zero. In the same time, the *Q* factor of the SBM in the single pattern structure is quite low (around 100) compared to those of the new modes. Thus, the original mode is not much perturbed since no other mode is introduced at its vicinity in the spectrum. Then, by increasing the magnitude of the perturbation, the *Q* factors of all the modes are decreased, and so their coupling with incident light increases, leading to broader peaks and thus more absorption [9]. It can also be observed that the resonant wavelength of the each mode changes with the magnitude of the perturbation. Combined with the broadening of the absorption peaks, this leads to an overlap of the modes. For high magnitude perturbations, the modes are so perturbed that the modal analysis turns out to be inaccurate. It can only be noticed that the significant modification in the wavelength range of 600 - 650 nm is due to an increased transmission into the cluster of holes that appears in the super cell, inducing a channelling effect [25].

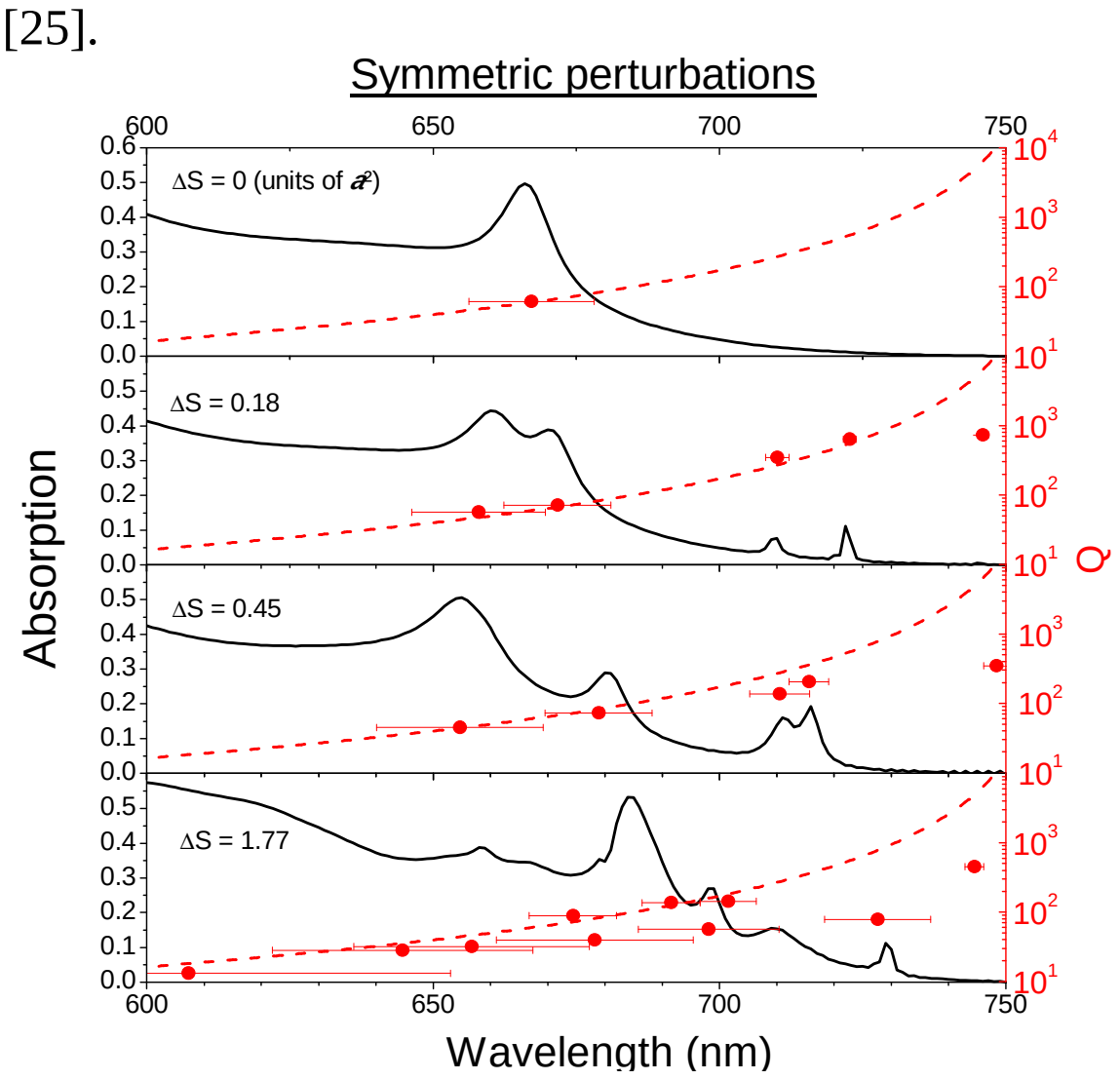

Fig.4 : Absorption spectra for various magnitudes of symmetric perturbation, as depicted in Fig.2B, in black line, left axis. *Q* factors of the modes, in red dots, right axis. The red dashed line depicts the critical coupling values.

It should be specified that the other type of symmetric perturbation, where the 4 holes located at the middle of the super cell sides are moved towards its centre, gives similar results which are not presented here.

As for the symmetric perturbation, additional modes appear for the non-symmetric and pseudo-disordered perturbations, in an even larger number. Moreover, the pseudo-disordered case has two specific properties: it is statistically isotropic, and by moving all the holes simultaneously, one can reach higher perturbation magnitudes compared to non symmetric perturbations.

To go further, the corresponding integrated absorption and the short circuit current density Jsc are plotted in Fig.6 for the three perturbations studied.

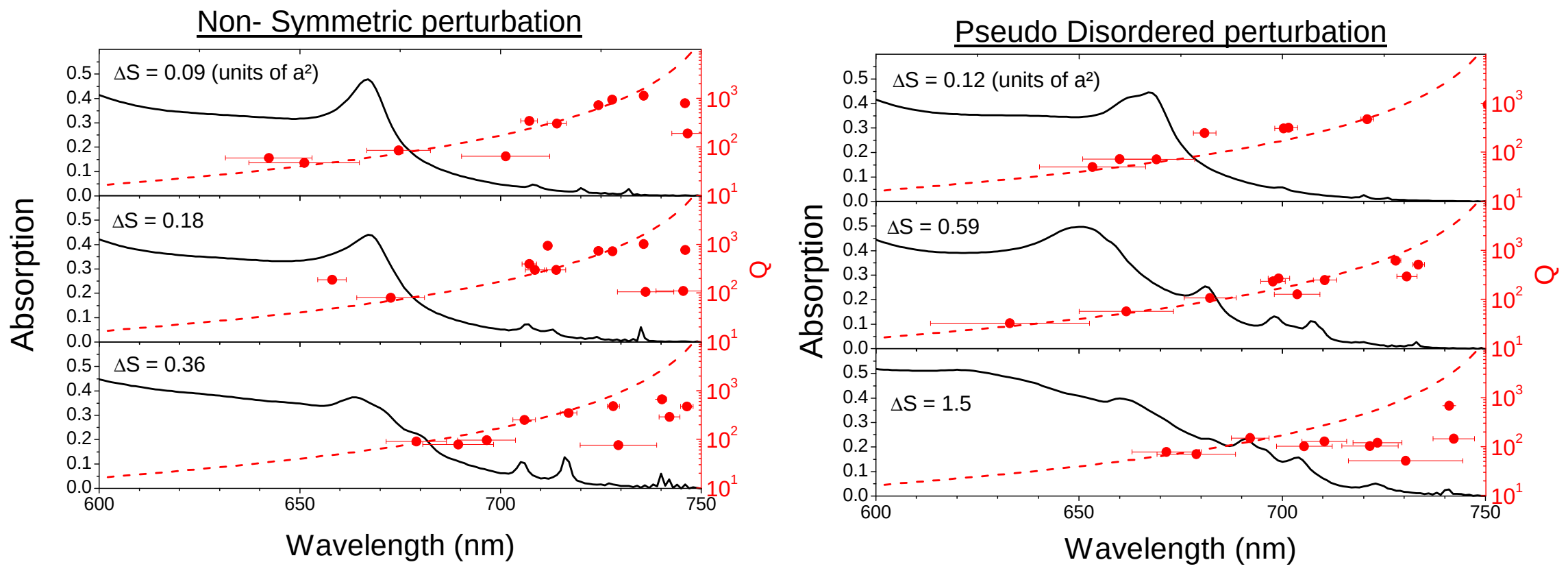

Fig.5 : Black lines: absorption spectra for the *x* polarization and for various magnitudes of non-symmetric perturbation, as depicted in Fig.2C (left) and pseudo-disordered perturbation, as depicted in Fig.2D, (right). Red dots: *Q* factor of the modes. The red dashed lines depict the critical coupling values.

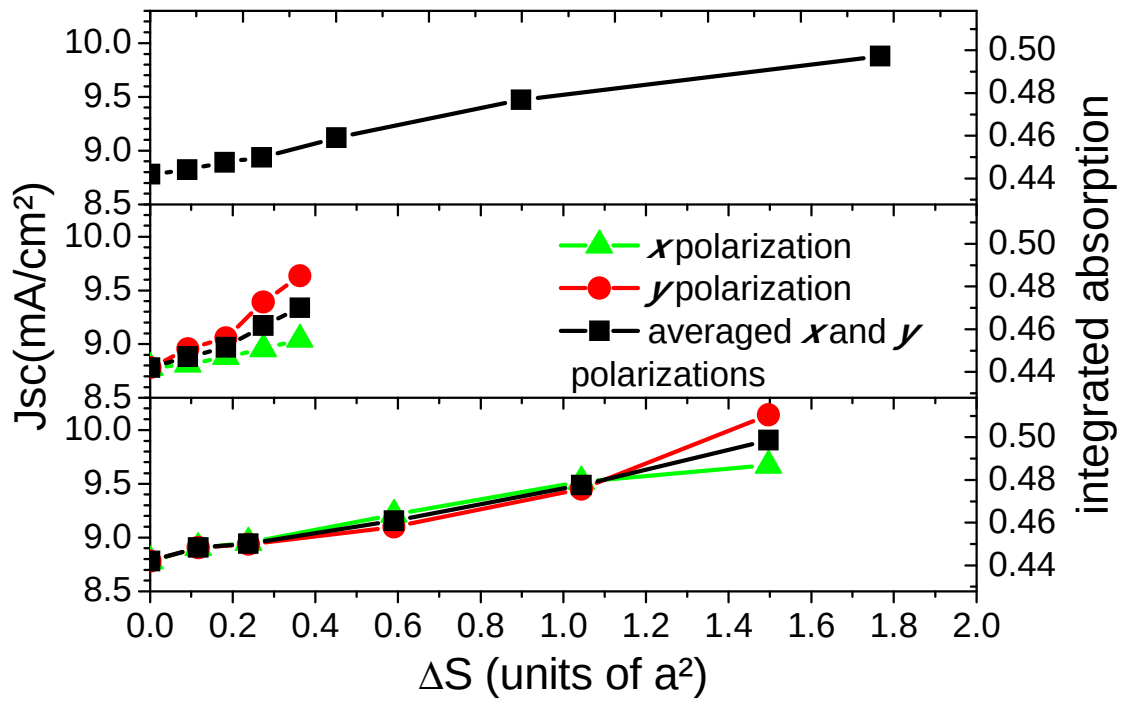

Fig.6 : Short-circuit current density and integrated absorption calculated for different magnitudes of the perturbation and for the 3 different kinds of perturbation: (top) symmetric, (middle) non-symmetric, and (bottom) pseudo-disordered perturbations.

For a small perturbation ΔS, the integrated absorption enhancement is larger for the non-symmetric perturbation than for the symmetric perturbation, since more modes are introduced. However, it can be noticed that for the non-symmetric perturbation, avoiding the overlap between two adjacent holes leads to a smaller maximum magnitude of the perturbation and therefore, one cannot overcome the integrated absorption obtained for the symmetric perturbation. In the case of pseudo-disorder, the enhancement versus perturbation is very similar to the symmetric case even if more modes are involved.

One can observe that the absorption is significantly enhanced when the perturbation is introduced. The maximum absolute increase of integrated absorption is about:

- 5.5%, for the symmetric case,
- 1.5% for x polarization, 4.5% for y polarization, and so a 3% mean value for the non-symmetric case. The modes appear to be more sensitive to a displacement of the hole in a direction perpendicular to the polarization of the light,
- 4.5% for x polarization, 7% for y polarization, and so a 5.75% mean value for the pseudo-disordered case. Additionally, due to the more isotropic pattern, there is less difference between the two polarization cases than for the previous perturbation.

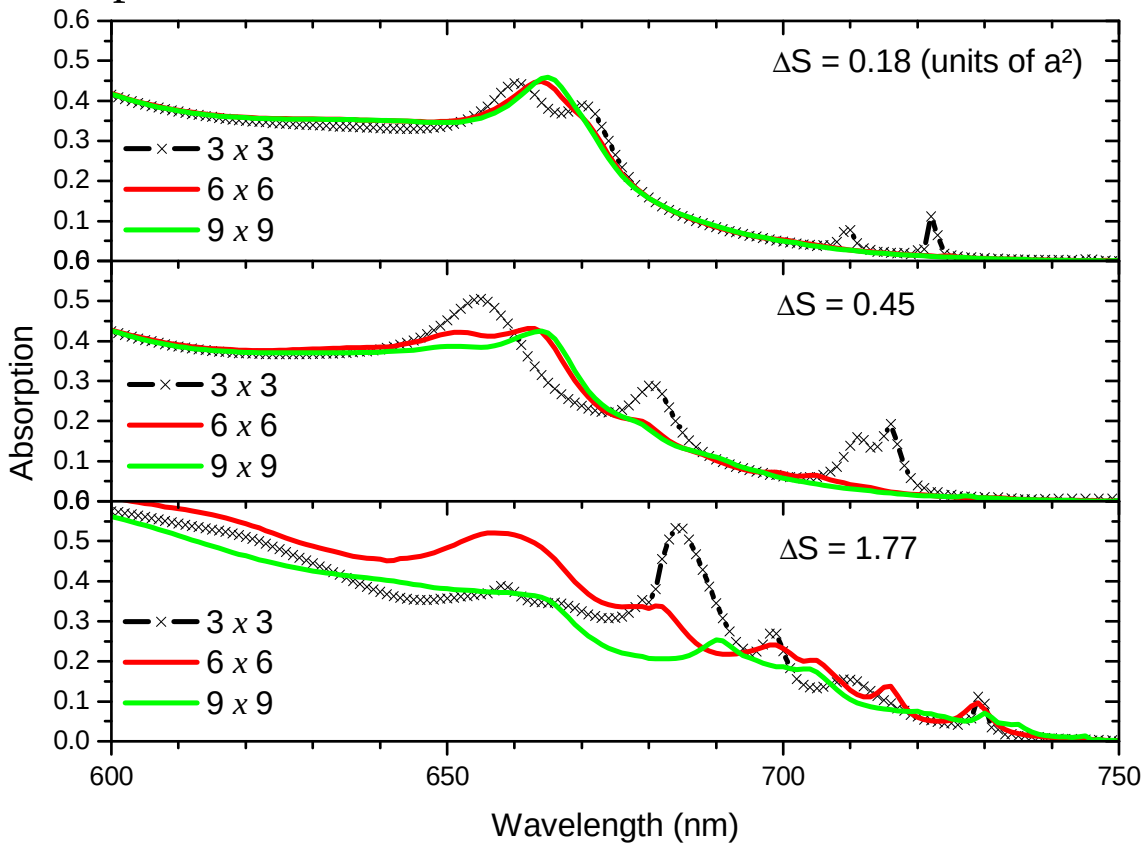

Fig.7 : Absorption spectra for different magnitudes of perturbation for 3 sizes of the super cell: 3 x 3 in black lines with crosses, 6 x 6 in red, and 9 x 9 in green (grey).

Let us now consider the case of larger symmetric super cells. Their absorption spectra are shown in Fig.7.

As the symmetric perturbation increases, the main absorption peak around 666 nm is smoothened for the larger super cells. At large perturbations, this leads to an integrated absorption for the 9 x 9 super cell lower than for the 3 x 3 super cell, the 6 x 6 being the best configuration in terms of integrated absorption, as can be seen in Fig 8. Consequently, it appears to be an optimal size of the super cell for the optimization of the integrated absorption.

To draw some preliminary conclusions about these perturbations, their larger absorption when compared to the single pattern case can be attributed to additional modes for small ΔS, and to a better coupling of those modes with the incident light leading to lower Q factors [9].

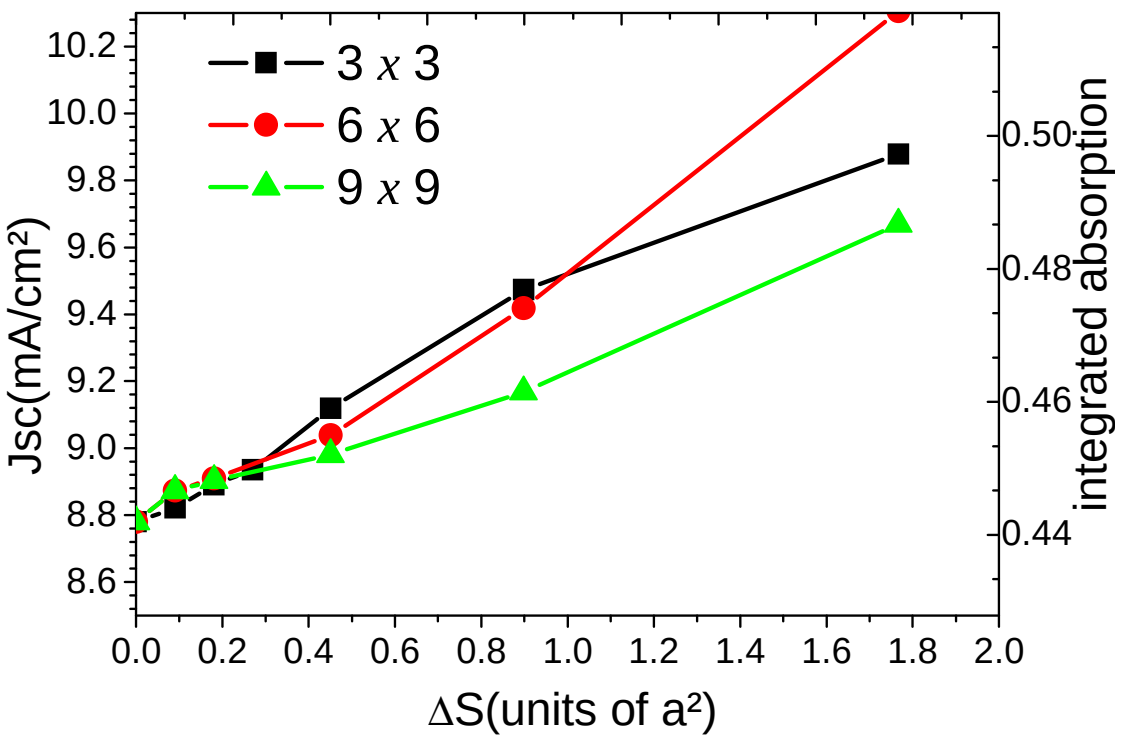

Fig.8 : Short-circuit current density and integrated absorption versus magnitude of the perturbation for the 3 sizes of the super cell: 3 x 3 in black, 6 x 6 in red, and 9 x 9 in green.

# V. ANALYSIS OF THE SPECTRAL DENSITY OF MODES IN THE RECIPROCAL SPACE

This section aims at providing an in-depth analysis of the origin and of the interest of the additional modes generated in the multiple pattern super cells. Indeed, in addition to the broadening of the absorption peaks, we will show that the increase of the spectral density of modes provides an essential contribution to the absorption enhancement.

We will first consider the band diagram of the single pattern unit cell PhC membrane. Next, we will discuss on the effects of the introduction of multiple patterns in a *n x n* super cell on the reciprocal lattice, on the basis of theoretical arguments. Then, the 3 *x* 3 symmetric case, owning the lowest number of additional modes, will be analysed, followed by the non-symmetric case. Next, we will relate the mode quality factor decrease to the magnitude of the perturbation. Finally, the influence of the size of the super cell on the absorption properties will be discussed.

The band diagram of the single pattern unit cell PhC membrane is plotted in Fig.9 [26], together with the light cones with respect to glass and to air.

It can be checked that the increased absorption around 666 nm in the absorption spectrum (Fig.3) is due to a single SBM in Γsp, where Γ denotes here the vertical direction (in plane wave vector k// = 0) and the subscript 'sp' stands for 'single pattern'. This mode (purple dot) is symmetric with respect to the in-plane crystallographic directions, whereas two non coupling (purple crosses) and anti-symmetric modes also exist in the same spectral range. The large number of modes that can be seen at 2Xsp/3 and 2Msp/3 may be involved in the further analysis of the multiple pattern super cell PhC.

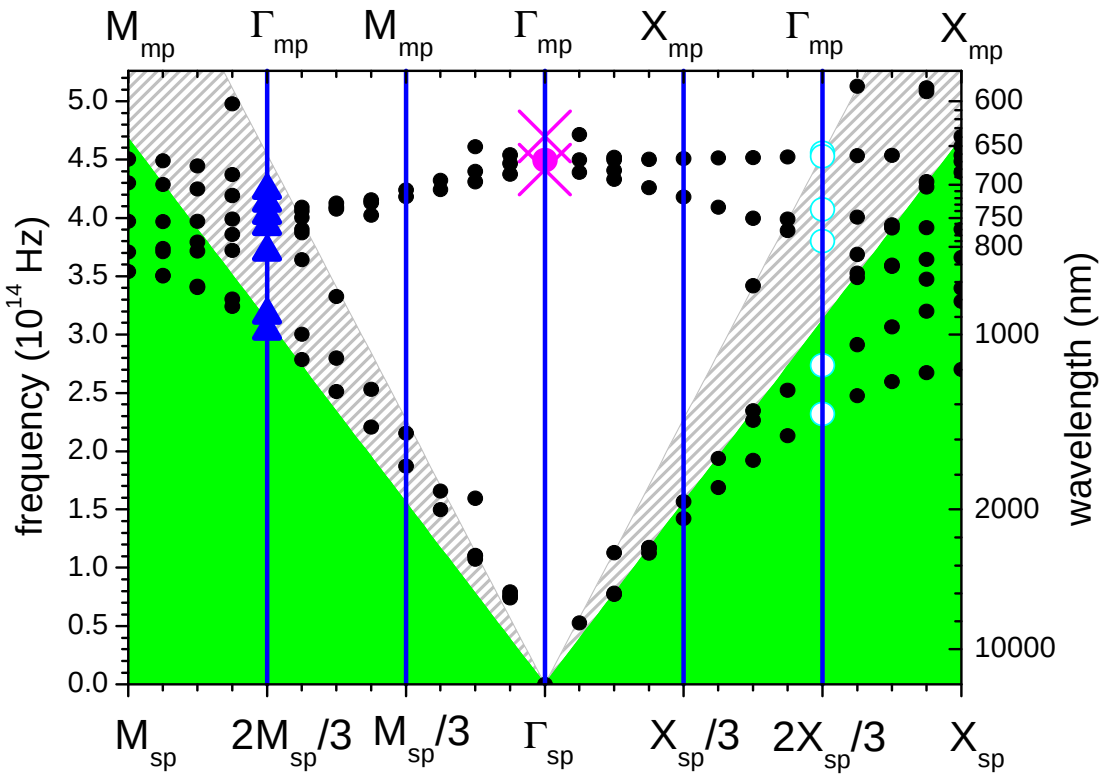

Fig.9 : Band diagram of the single pattern unit cell 2D PhC membrane. At $\Gamma_{sp}$ point, the coupling mode enhancing the absorption is represented by a purple dot and the non coupling modes by purple crosses. In $2M_{sp}/3$ the modes are depicted by blue triangles, in $2X_{sp}/3$ the modes are depicted by light blue hollow circles. The light cone corresponding to the substrate is in light grey hatched pattern, the one corresponding to the air is in plain green.

## A. Reciprocal lattice analysis in a *n x n* symmetric multiple patterns super cell

To better understand the origin of the additional modes coupled under normal incidence when multiple patterns are introduced in the super cell, we theoretically describe the evolution of its reciprocal space.

The general representation of the reciprocal lattice space of a 2D square periodic structure is reported in Fig.10 (left). The Γ *k*-points, located at the nodes $\left(l.\vec{G}_1, m.\vec{G}_2\right)$ of the lattice, with $\vec{G}_1$ and $\vec{G}_2$ the elementary vectors and *l* and *m* being two even integers, are the main *k*-points of interest in our study, since they are the only *k*-points that can be addressed by a plane wave under normal incidence, thanks to the diffraction mechanism. More precisely, it can be shown (see Appendix for a detailed demonstration) that at each set of 4 Γ of these points having the same $|k_{//}|$ (so located on a circle centred at (0, 0) and with a radius $|k_{//}|$) corresponds one mode that can couple to such a plane wave, and thus one peak in the absorption spectrum of the PhC membrane.

One can thus superimpose the *n.a*- periodic multiple pattern (subscript 'mp') structure lattice space with the one of the previous single pattern structure (see Fig.10 (right)). One conclusion and one comment can be driven from this schematic.

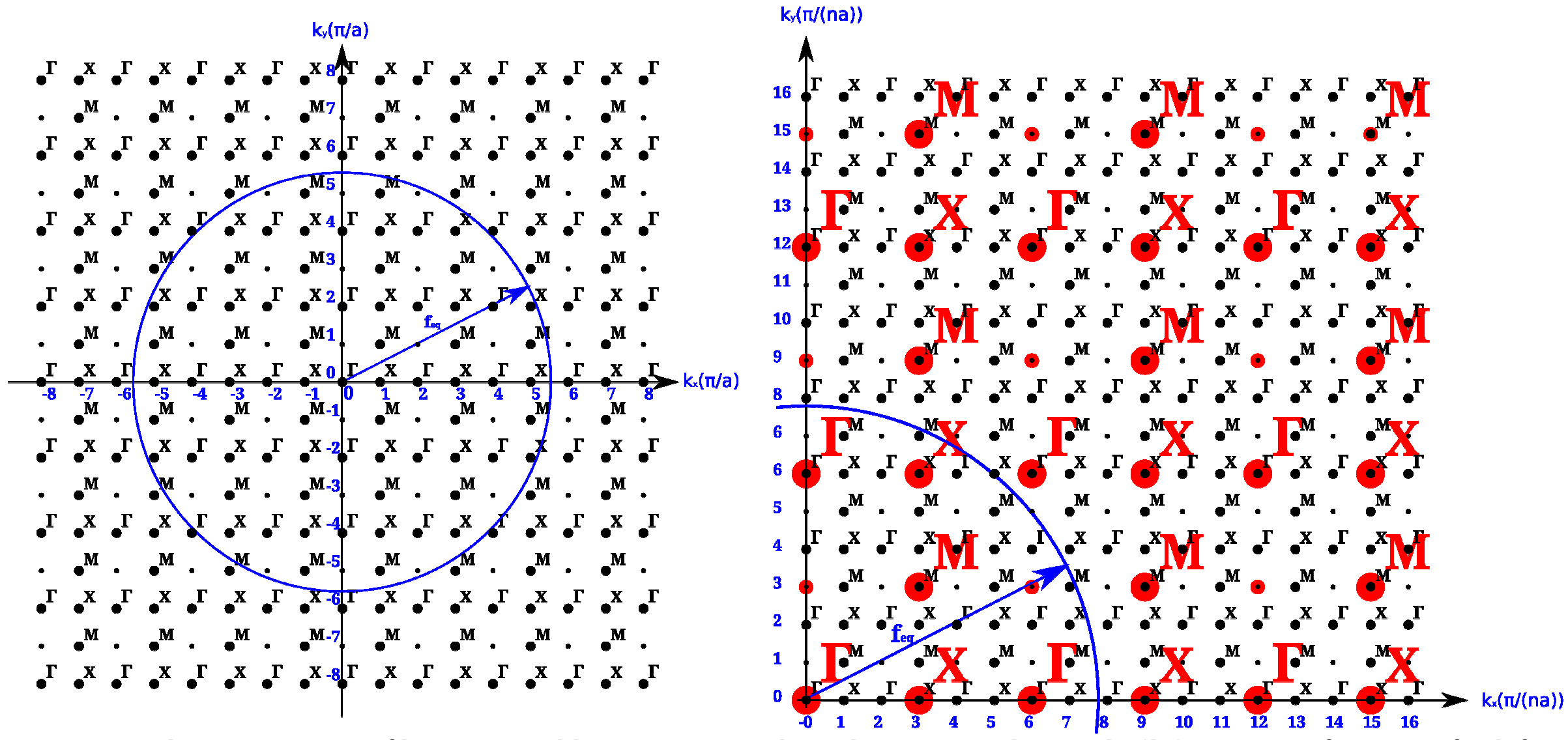

*Fig.10. General representation of he reciprocal lattice space, with circle corresponding to the $|k_{//}|$ at a given frequency $f_{eq}$ (left); Superimposition of two reciprocal lattices, corresponding to a a periodic PhC (red, big dots) and a n.a periodic PhC (black, small dots). n = 3 on this example (right).*

Firstly, thanks to this superimposition, it can be seen that numerous $\Gamma_{mp}$ points are added, compared to the number of $\Gamma_{sp}$ points of the single pattern unit cell PhC reciprocal lattice. Based on the demonstration developed in Appendix, it can be derived that the number of modes that can couple to a plane wave under normal incidence (thanks to their double degeneracy [18]) in a multiple pattern super cell structure is $n^2$ times the number of the similar kind of modes existing in the single pattern unit cell structure, provided that the spectral range of interest is broad enough. This leads to $n^2$ times more peaks in the absorption spectrum over the same spectral range.

Moreover, it can be noticed that the additional modes arise from modes at $k_{sp}$ points of different symmetries ('band folding' mechanism). It should be thus possible to predict where the modes will appear in the absorption spectrum of the multiple pattern super cell PhC under normal incidence by only simulating the case of the single pattern unit cell under oblique incidence, and then by analyzing the symmetry of the modes but in the multiple pattern geometry to predict if their coupling with a plane wave is allowed or forbidden.

## B. Spectral density of modes in a nxn non-symmetric multiple patterns super cell

Similar developments can be achieved in the non-symmetric case. However, since the symmetries are lost, this breaks the previously established degeneracy, and thus introduces two times more additional modes. Finally, all other modes that could not couple to a plane wave under normal incidence for symmetry reasons (twice the number of symmetric modes [without degeneracy, see Appendix, Table 3]) can now be coupled. Given these two ratios, for small perturbations, about four times more modes than for the symmetric perturbation should thus be coupled.

## C. Example of 3 x 3 symmetric perturbations

The previous superimposition method is applied for the case $n = 3$. Considering the band diagram of the single pattern unit cell PhC, the modes involved in the absorption over the spectral range of interest are around the $k_{sp}$ points of the reciprocal space having $|k_{//}| = 2\pi/a$. Given the previous demonstration, the resulting possible modes can be found at the various $k_{mp}$ points having a $|k_{//}|$ reported in Table 1 : .

Fig.11 clearly illustrates the folding mechanism consequences by identifying the SBM responsible for the larger absorption.

It can be seen that the numerical modal analysis exhibits 5 additional modes, among all the possibilities reported in Table 1 : . However the absorption is already large enough at wavelengths smaller than the one of the single absorption peak of the single pattern unit cell PhC. This annihilates all the new modes in the multiple pattern super cell PhC at lower wavelengths. Thus, only some of the additional modes at larger wavelengths will significantly contribute to the absorption.

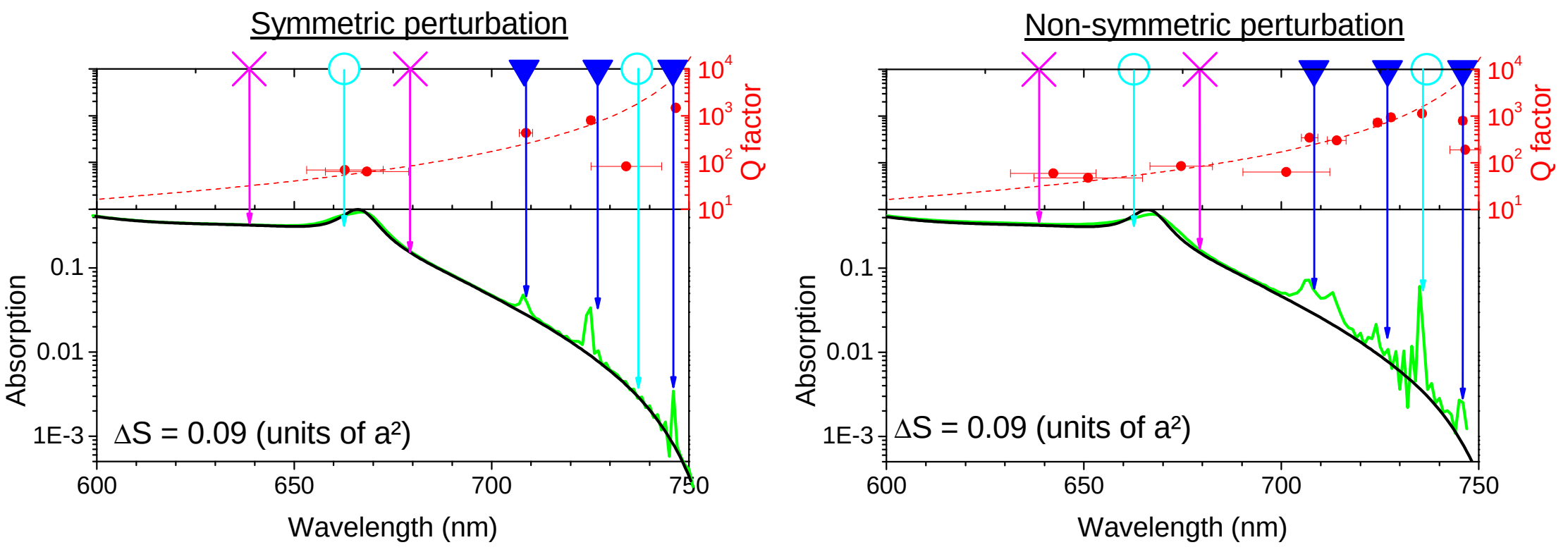

Fig.11 : (bottom) Absorption of the PhC membrane without perturbation (in black) together with the symmetric perturbation for $\Delta S = 0.09a^2$ (in green (grey)), for a symmetric (left) or non-symmetric (right) perturbation; (top) $Q$ factor of the SBM of the perturbed sample, in red, right axis. The symbols on top and the associated arrows depict the modes resulting from the folding mechanism applied to the original band diagram (Fig.9). Purple crosses are non coupling mode in Γ point, light blue circles are SBM in $2X_{sp}/3$ and dark blue triangles are SBM in $2M_{sp}/3$.

| $\lvert k_{//mp}\rvert$ (in $\pi/na$ units) | 2 | $2\sqrt{2}$ | 4 | $2\sqrt{5}$ | $4\sqrt{2}$ | 6 (=$k_{sp}$) | $2\sqrt{10}$ | $2\sqrt{13}$ | $6\sqrt{2}$ |
|---|---|---|---|---|---|---|---|---|---|
| Theoretical number of peaks (doubly degenerate modes) contributing to the absorption in the spectral range of interest | 1 | 1 | 1 | 2 | 1 | 1 | 2 | 2 | 1 |

Table 1 : Number of modes that can couple to an incident plane wave under normal incidence for various $\lvert k_{mp}\rvert$ around a given $k_{sp}$, according to group theory considerations (see Appendix).

More precisely, given the small perturbation applied in our example, the frequencies of the additional modes should correspond to the frequencies of the modes at the corresponding $k_{sp}$ points of the single pattern unit cell PhC reciprocal lattice. The identification shows the three modes from $2M_{sp}/3$ (at 708 nm, 727 nm and 746 nm) and 2 out of the 3 SBM from $2X_{sp}/3$ (at 663 nm and 737 nm), but only the SBM from $2M_{sp}/3$ clearly generate peaks in the spectrum. Indeed, the first SBM at $2X_{sp}/3$ is spectrally too close to the one in $\Gamma_{sp}$ and the other one (at 737 nm) exhibits relatively low $Q$ compared to $Q_c$ (86 to be compared to 2000) at $2X_{sp}/3$, and so is too broad to be clearly visible. It has also to be noticed that the $Q$ factor of the mode at the highest wavelength in $2X_{sp}/3$ is very low even without perturbation due to its losses in the substrate (this SBM, as the others, stands below the air cone but above the glass cone and so exhibits radiative losses), explaining why the harmonic inversion analysis can detect it whereas it cannot be seen on the spectrum.

Due to the symmetry conservation, the two other SBM at the Γ point that cannot couple to a plane wave under normal incidence without perturbation, can still not couple using this perturbation (purple crosses in Fig.11).

## D. Case of non-symmetric perturbations: mode splitting and mode coupling

As can be seen in Fig.11, the folded modes are now split into two, thanks to the symmetry breaking of the PhC. Moreover, in addition to these SBM, the other SBM at the Γ point that could not couple with symmetric perturbation can now be coupled, as it is the case for the wavelength close to 680 nm.

As a conclusion on the ‘band folding’ mechanisms at the Γ point, additional SBM result from the $k$ point folding in the reciprocal lattice of the single pattern unit cell PhC, thanks to the momentum introduced by the perturbation. In the non-symmetric case, this effect is even twice more important due to the splitting of the modes, resulting from the symmetry breaking. Moreover, SBM at the Γ point that could not be coupled without perturbation for symmetry reasons can now be coupled and contribute to the absorption.

These qualitative conclusions for the non symmetric case can be applied to the pseudo-disordered one.

Thanks to these mechanisms, we also address the Fourier spectrum filling described in other studies [8, 11, 10]. The new modes appearing in the Fourier spectrum are the ones resulting from the above-mentioned folding and are originating from $k$-points depending on the size of the super cell.

## E. Influence of the perturbation on the *Q* factors of the modes

In addition to the increase of the spectral density of modes [7, 8], the integrated absorption of each peak can be increased thanks to a reasonable decrease of their *Q* factor, as was analysed by *Chutinan et al.* [14], observed by *Oskooï et al.* [16] and modelled by *Gomard et al.* [9]. Thus, it is also important to compute the *Q* factors of the modes in the perturbed PhC as a function of the magnitude of the perturbation. This point is investigated in Fig.12. The *Q* factors of the SBM at around 666 nm of the single (unperturbed) pattern unit cell PhC membrane at the Γ point and those at around 745 nm of the SBM at the largest wavelengths in the spectral range considered (coming from the folding of the $2M_{sp}/3$ *k* point) are tracked.

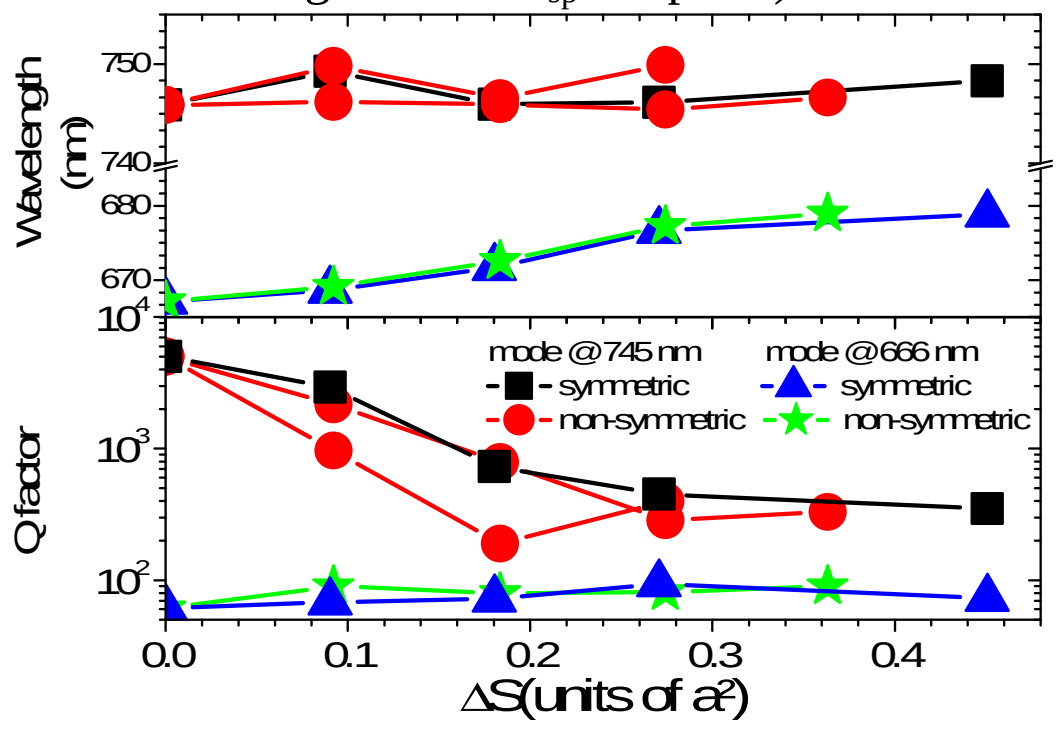

Fig.12 : Influence of the magnitude perturbation on the resonant wavelength (top) and on the quality factor (bottom) of the degenerate modes previously in Γ (around 666 nm) for the symmetric (black square marks) and non-symmetric (red dots) perturbations and in $2M_{sp}/3$ (around 745 nm) for the symmetric (blue triangles) and non-symmetric (green stars) perturbations.

It can then be seen in Fig.12 that the *Q* factor of the SBM is almost not affected by the perturbation. Actually, the resonant wavelengths remain almost constant for both modes. Moreover, the mode losses for the single pattern unit cell PhC membrane are already relatively high and the introduced perturbations will not significantly increase those losses, leading to quite unchanged *Q* factors. However, for the SBM at the largest wavelength, the initial *Q* factor of the mode at $2M_{sp}/3$ is high (5000, due to really small material absorption and weak coupling with the incident light at this particular angle). Consequently, it is much more sensitive to the introduction of additional losses thanks to the perturbation. The decrease of *Q* factors by a decade for both symmetric and non symmetric perturbations, and for both modes coming from the splitting in the non symmetric perturbation, strongly contributes to the integrated absorption increase.

## F. Effect of the size of the super cell

As shown previously, the size of the super cell will play a major role in controlling the number of modes that can be added by the perturbation. Because there are a lot of additional SBM in the 6 *x* 6 and 9 *x* 9 super cell compared to the 3 *x* 3 one (the number of additional modes follows a quadratic law), all the SBM overlap spectrally.

This leads to a weaker coupling into each SBM when the modal density is high, thus shifting the coupling condition from its optimum. This can be seen as an increase of the backward diffraction induced by the larger number of diffracted order with the larger period.

As was shown numerically in Fig.8, it then appears that there is an optimal size, which consists in a compromise between the increase of the spectral density, and the decrease of the coupling. However, this optimum strongly depends not only on the size, but also on the nature of the perturbation considered and as such, cannot be predicted easily.

Based on of this spectral density of modes analysis using the superimposition of the reciprocal lattices, it can be concluded that by introducing a perturbation in a PhC membrane, new radiative losses that decrease the *Q* factor of the SBM are created. This effect can be used to enhance the absorption in solar cells or detectors since a coupling to the incident wave larger than the losses is the most appropriate configuration for a broadband absorption. In addition, the limit of the *Q* factor of the additional SBM for small perturbations can be predicted. This gives an indication on the potential enhancement of the absorption in a given material.

# VI. ABSORPTION IMPROVEMENT BY PSEUDO DISORDER IN AN OPTIMIZED PHOTONIC CRYSTAL MEMBRANE

In the previous sections, the case of a single pattern unit cell PhC membrane exhibiting only one SBM of interest was considered. It allowed us to identify the mechanisms responsible for the increase of the modal density and the mode width broadening, possibly leading to an increase of the integrated absorption. The relevancy of our approach is now demonstrated by checking that such a patterning can lead to higher integrated absorptions than for an optimized but single pattern PhC membrane.

Using the same membrane as in Fig.1, and keeping the previous layer thickness and hole depth, the overall absorption (calculated between 400 nm and 750 nm) can be maximized for *a* = 345 nm and *ff* = 50% according to a particle swarm optimization algorithm [19]. These single pattern unit cell PhC membrane parameters allow to harvest 58% of the incoming photons, leading to a maximal short circuit current density $J_{sc}$ of 11.5 mA/cm². This value is already very close to the one obtained in the Lambertian regime (integrated absorption of 59.3% and $J_{sc}$ of 11.8 mA/cm²), despite the lack of an anti reflective top coating like ITO. However, one can see that two spectral regions located at around 620 nm and above 720 nm are free of modes. To increase even more the absorption by filling these gaps, we introduce a pseudo-disordered perturbation (see Fig.2D) on a 3 *x* 3 super cell with a magnitude of $\Delta S = 0.6a^2$. The absorption spectra (for polarization along *x* and *y* and the average) are compared to the one obtained for the layer without perturbation but optimised in Fig.13.

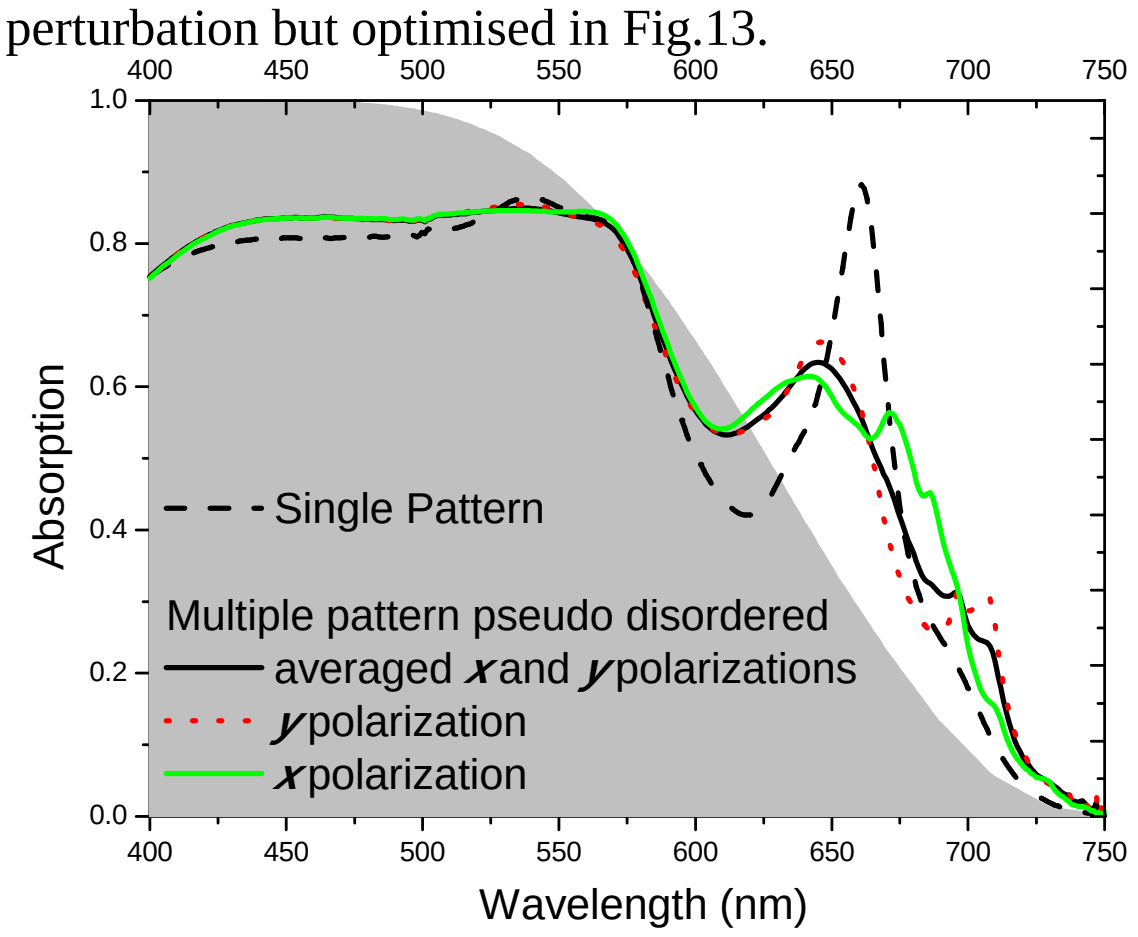

Fig.13 : Absorption spectrum for the optimised single pattern in dashed black ($a$=345 nm, *ff*= 50%) compared to the pseudo disordered structure (average of the two polarizations in black, *y* polarization in dotted red, *x* polarization in green (grey)). The absorption spectrum obtained with a Lambertian model is also depicted by the grey background

The main differences appear in the low absorbing region (wavelength > 600 nm) where both structures exhibit a higher absorption than the one calculated with the Lambertian model [11]. The pseudo-disordered layer presents smoother and broader absorption peaks. Additionally, at short wavelengths, where a-Si:H is highly absorbing, both structures absorb less than the Lambertian model, but show a similar behaviour. This is again due to the fact that there is no anti reflection layer. The global integrated absorption of the multi patterned structure is 60% ($J_{sc}$ = 11.9 mA/cm²), and therefore is higher than the unperturbed structure, and even slightly higher than the Lambertian model.

Let us apply our previous analysis and focus on the absorption spectra for both polarizations in the low absorbing region. Additional modes appear in the pseudo-disordered structure and the decrease of the $Q$ factors leads to a broadening of the modes already existing in the single pattern structure. These observations follow exactly the main conclusions established in the previous sections. It is also important to note that this was observed for several pseudo-disordered patterns with about the same ΔS. However, this absorption enhancement is observed only for ΔS around $0.6a^2$. Such an optimal range of the magnitude of perturbation has already been observed [10, 16].

This example demonstrates that the controlled introduction of pseudo-disorder enables to overcome the integrated absorption obtained for an already optimised single pattern unit cell 2D PhC membrane, as a consequence of additional modes and more favourable $Q$ factor values.

# VII. CONCLUSION

The enhancement of the absorption allowed by a $n$ *x* $n$ perturbation of a 2D square PhC in an a-Si:H membrane, close to the bandgap of this material, was studied in several ways. Firstly, several displacement perturbations in a 3 *x* 3 patterns 2D square lattice PhC were introduced, with various symmetry properties. All these perturbations enhanced the integrated absorption of a non optimized, single mode 2D PhC membrane. The optical gains computed are attributed both to the about $n^2$ times more SBM resulting from the folding of the reciprocal space, and to the larger losses of these modes leading to better coupling conditions with the incident plane wave. Among the perturbations studied, the non-symmetric ones are the most promising since they exhibit $4n^2$ more SBM than in the single pattern unit cell PhC, instead of $n^2$ times more modes with a symmetric perturbation.

The symmetric perturbation was then studied for several sizes of super cells (3 *x* 3, 6 *x* 6 and 9 *x* 9), and we observed an optimal size corresponding to the 6 *x* 6 case. It corresponds to a compromise between a larger spectral density of modes and a weaker coupling of these modes with the plane wave under normal incidence.

All these effects allow an absorption enhancement of an already optimized single pattern unit cell PhC membrane using simply a non optimal 3 *x* 3 pseudo-disordered super cell PhC. The integrated absorption computed over the 400 nm – 750 nm range is increased from 58 % to 60 %. This should be further increased by a further optimization the pseudo-disordered super cell (mainly the size of the perturbation). In addition, the number of patterns of the super cell indicates the number and the approximate frequencies of the modes resulting from the band folding, whereas the $Q$ factors of these resulting SBM can be tuned thanks to the magnitude of the perturbation. It is thus quite straightforward to optimise the absorption at a specific wavelength of the spectrum. This method could be especially applied to reach really low coupling, or high $Q$ factor, in order to address really low absorption at a specific wavelength, or in other point of view, to control the output losses of SBM for other applications such as lasers or sensors. Finally, it can be noticed that such structures could be judiciously fabricated using nanoimprint lithography, since the resolution is large enough and the throughput rates already reach an industrial level [27, 28].

## VIII. ACKNOWLEDGMENT

The authors are grateful to PICM laboratory, Palaiseau, for providing the optical indices, to Xavier Letartre, Taha Benyattou and Pierre Viktorovitch for discussions, to Pierre Clare for helping to the mode counting, and to the referee for helping us to improve the paper.

This work was supported by the European Community (FP7 project PhotoNvoltaics N°309127) and by Orange Labs Networks (contract 0050012310-A09221).

## IX. REFERENCES

[1] S. John, Nature Materials **11**, 997 (2012).
[2] S. Mallick, N. Sergeant, M. Agrawal, J. Lee, and P. Peumans, MRS BULLETIN **36**, 453 (2011).
[3] N. Reich, W. van Sark, and W. Turkenburg, Renewable Energy **36**, 642 (2011).
[4] W.-C. Lai, S. Chakravarty, X. Wang, C. Lin, and R. T. Chen, Applied Physics Letters **98**, 023304 (2011).
[5] R. Peretti, G. Gomard, C. Seassal, X. Letartre, and E. Drouard, Journal of Applied Physics **111**, 123114 (2012).
[6] R. Peretti, G. Gomard, C. Seassal, X. Letartre, and E. Drouard, Tailoring the absorption in a photonic crystal membrane: A modal approach, in *Proceeding SPIE*, edited by H. R. Mguez, S. G. Romanov, L. C. Andreani, and C. Seassal Vol. 8425, p. 84250Q, SPIE, 2012.
[7] Z. Yu, A. Raman, and S. Fan, Applied Physics A: Materials Science & Processing **105**, 329 (2010), 10.1007/s00339-011-6617-4.
[8] Z. Yu, A. Raman, and S. Fan, Opt. Express **18**, A366 (2010).
[9] G. Gomard, R. Peretti, E. Drouard, X. Meng, and C. Seassal, Opt. Express **21**, A515 (2013).
[10] E. R. Martins, J. Li, Y. Liu, J. Zhou, and T. F. Krauss, Phys. Rev. B **86**, 041404 (2012).
[11] L. Andreani, A. Bozzola, P. Kowalczewski, and M. Liscidini, Towards the lambertian limit in thin film silicon solar cells with photonic structures, in *27th European Photovoltaic Solar Energy Conference and Exhibition: Material Studies, New Concepts and Ultra-High Efficiency Fundamental Material Studies*, 2012.
[12] F. Pratesi, M. Burresi, F. Riboli, K. Vynck, and D. S. Wiersma, Optics Express **21**, A460 (2013).
[13] C. Lin, L. J. Martínez, and M. L. Povinelli, Opt. Express **21**, A872 (2013).
[14] A. Chutinan and S. John, Phys. Rev. A **78**, 023825 (2008).
[15] K. Vynck, M. Burresi, F. Riboli, and D. Wiersma, Nature Materials **11**, 1017?1022 (2012).
[16] A. Oskooi *et al.*, Applied Physics Letters **100**, 181110 (2012).
[17] A. Bozzola, M. Liscidini, and L. C. Andreani, Opt. Express **20**, A224 (2012).
[18] K. Sakoda, *Optical properties of photonic crystals* Vol. 80 (Springer series in optical science, 2001).
[19] http://www.lumerical.com/.
[20] E. D. Palik, *Handbook of Optical Constants of Solids: Index* Vol. 3 (Academic press, 1998).
[21] G. Gomard *et al.*, Journal of Applied Physics **108**, 123102 (2010).
[22] V. Mandelshtam, Progress in Nuclear Magnetic Resonance Spectroscopy **38**, 159 (2001).
[23] A. S. for Testing and M. Philadelphia, (2010).
[24] Y. Park *et al.*, Opt. Express **17**, 14312 (2009).
[25] G. Gomard *et al.*, not published (2013).
[26] G. Gomard *et al.*, Journal of Optics **14**, 024011 (2012).
[27] M. Malloy and L. C. Litt, Journal of Micro/Nanolithography, MEMS, and MOEMS **10**, 032001 (2011).
[28] F. Lenzmann, J. Salpakari, A. Weeber, and C. Olson, Light-trapping in solar cells by photonic nanostructures: the need for benchmarking and fabrication assessments, by: Publication date: ECN Solar Energy 16-6-2013, 2013.
[29] J. Rosenberg, R. V. Shenoi, S. Krishna, and O. Painter, Opt. Express **18**, 3672 (2010).
[30] H. Sakamoto, T. Uno, and T. Arima, Classification of degenerate and non-degenerate modes of photonic crystals by group theory in fdtd analysis, in *Electromagnetic Theory (EMTS), Proceedings of 2013 URSI International Symposium on*, pp. 716–718, IEEE, 2013.

## X. APPENDICES

### A. Optical indices

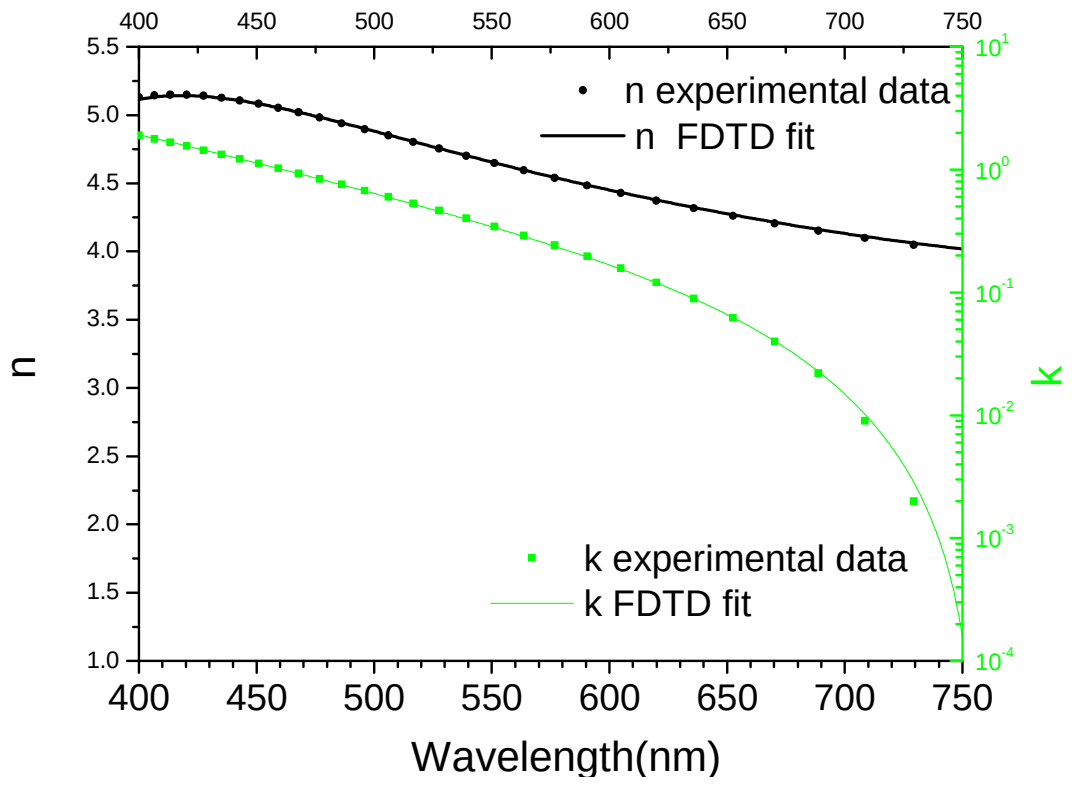

Fig.14 : Measured and fitted optical indices of the a-Si:H layer

The optical indices of a-Si:H used in the simulations have been fitted from ellipsometric measurements. The good agreement between measured and fitted data for FDTD simulations can be checked in Fig.14.

Without loss of generality, the silica has been considered as transparent, with a refractive index of 1.5.

## B. FDTD simulation conditions

The same refinements were used for the non-uniform spatial mesh for all the FDTD simulations, leading to an around 3.5 nm mesh step size. The temporal steps and number of iterations remain constant, corresponding to a simulated time of 1000 fs, to allow an accurate modal analysis. Absorption spectra have been obtained using plane waves under normal incidence, whereas resonance frequencies and Q factors have been obtained using dipole source inside the structure apart from high symmetry points. All these conditions enable a reliable comparison between the various spectral and integrated absorption numerical data.

## C. Irreducible representations and number of modes at the various Γ points

| Possible field map symmetry properties | $C_{4v}$ | E | $C_2$ | $2C_4$ | $2\sigma_v$ | $2\sigma_d$ |
|---|---|---|---|---|---|---|
| 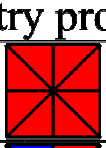 | *A1* | 1 | 1 | 1 | 1 | 1 |
| 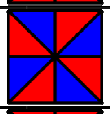 | *A2* | 1 | 1 | 1 | -1 | -1 |
| 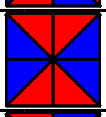 | *B1* | 1 | 1 | -1 | 1 | -1 |
| 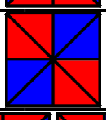 | *B2* | 1 | 1 | -1 | -1 | 1 |
| 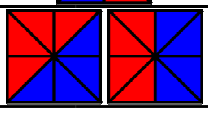 | *E* | 2 | -2 | 0 | 0 | 0 |

Table 2 : Symmetry representations of modes occurring at the Γ points of various |*k*| for a square lattice PhC. The first column gives examples of maps of the main field component along the *z* axis, for null azimuthal and radial order.

In this Appendix, we establish the density of modes in a square lattice PhC that can be coupled under a normally incident plane wave, i.e. at the Γ points.

As previously demonstrated in [18, 29, 30], all the modes at Γ points can be described using the symmetry operations shown in Table 2.

Only the modes corresponding to the representation E can be coupled to an incoming plane wave due to symmetry reasons [18]. Moreover, these modes are doubly degenerate.

To know what kind of modes arises from a given Γ point at a specific free space frequency, one have first to draw a circle of radius |k//| in π/na units (see Fig.10 (left)). Thus:

- when $k_{//}^2 = l^2$, the circle path crosses 4 Γ points on symmetry axes, 2 on the kx axes and 2 on the ky axes
- when $k_{//}^2 = 2l^2 = l^2 + m^2$ the circle path crosses 4 Γ points on symmetry axes, 2 on the kx + ky axes and 2 on the kx - ky axes
- when $k_{//}^2 = l^2 + m^2$ with l ≠ m and both ≠ 0, the circle path crosses 8 points out from any symmetric position.

For all these 3 cases one can follow and extend the demonstration given in [18]. Results are given in Table 3.

| Possible field map symmetry properties | C4v | $k_{//}^2=l^2$ | $k_{//}^2=2l^2$ | $k_{//}^2=l^2+m^2$ |
|---|---|---|---|---|
| 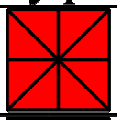 | *A1* | 1 | 1 | 1 |
| 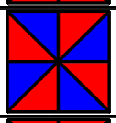 | *A2* | 0 | 0 | 1 |
| 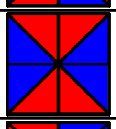 | *B1* | 1 | 0 | 1 |
| 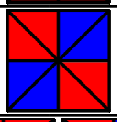 | *B2* | 0 | 1 | 1 |
| 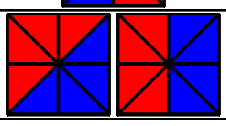 | *E* | 1 | 1 | 2 |

Table 3 : Modes arising from coupling of modes from Γ points lying at different symmetry positions. First column gives examples of maps of the main field component along the *z* axis, for null azimuthal and radial order.

To summarize this table, for each set of 4 modes ($k_{//}^2 = l^2$ or $k_{//}^2 = 2l^2$), then one E mode appears in the spectrum, and for each set of 8 modes ($k_{//}^2 = l^2 + m^2$) then 2 E modes appear in the spectrum.

Now, to count the number of modes that can couple with a normally incident plane wave, one just have to count each Γ point on Fig.10 (left) depending on the radius |k//| and divide the sum by 4. For radius large enough, the result is $\pi r^2$ where $r = ak_{//}/2\pi$ Thus, the number of mode for a k// large enough compared to 2π/a is $k_{//}^2 a^2 / 4\pi$.

Finally, let's compare the cases of single pattern and multiple pattern cell PhC. Provided the highest wavelength of the spectrum of interest is high enough, the spectral density of modes in the multiple pattern super cell PhC is $n^2$ times larger than one of the single pattern unit cell PhC. However this value remains true for n = 2 and n = 3 and for k// as small as 2π/a.